\documentclass[aps,twocolumn]{revtex4}

\usepackage[utf8]{inputenc}
\usepackage[T1]{fontenc}
\usepackage{graphicx}
\usepackage{lmodern} 

\usepackage[english]{babel}

\usepackage{xcolor} 

\usepackage{url}
\usepackage{hyperref}
\usepackage{textcomp}

\usepackage{amsmath,amsfonts,amssymb}
\usepackage{bm, bbold}

\usepackage{siunitx}
\usepackage[normalem]{ulem}

\newcommand{\ie}{i.\,e.}

\newcommand{\eg}{e.\,g.}
\newcommand{\imag}{\mathrm{i}}

\newcommand{\dif}{\mathrm{d}}
\newcommand{\diag}{\mathrm{diag}}

\newcommand{\std}{\mathrm{std}}
\newcommand{\mean}{\mathrm{mean}}

\newcommand{\nmed}{n_{\rm m}}

\begin{document}

\title{Extinction spectra of suspensions of microspheres: Determination of spectral refractive index and particle size distribution with nanometer accuracy}

\author{Jonas Gienger}
\email{jonas.gienger@ptb.de}
\author{Markus B\"{a}r}
\author{J\"{o}rg Neukammer}

\affiliation{Physikalisch-Technische Bundesanstalt (PTB), Abbestra\ss{}e 2--12, 10587 Berlin, Germany}

%
%

\date{Compiled \today}

%

\begin{abstract}
  A method is presented to infer simultaneously the wavelength-dependent real refractive index (RI) of the material of microspheres and  their size distribution from extinction
  measurements of particle suspensions. To derive the averaged spectral optical extinction cross section of the microspheres from such
  ensemble measurements, we determined the particle concentration by flow cytometry to an accuracy of typically 2\% and adjusted the particle concentration to ensure that
  perturbations due to multiple scattering are negligible. For analysis of the extinction spectra we employ Mie theory, 
  a series-expansion representation of the refractive index and nonlinear numerical optimization. 
  In contrast to other approaches, our method offers the advantage to simultaneously  determine size, 
  size distribution and spectral refractive index of ensembles of microparticles including uncertainty estimation.
\end{abstract}



\maketitle


\begin{figure*}[ht]
 \centering 
 \includegraphics[width=0.667\textwidth]{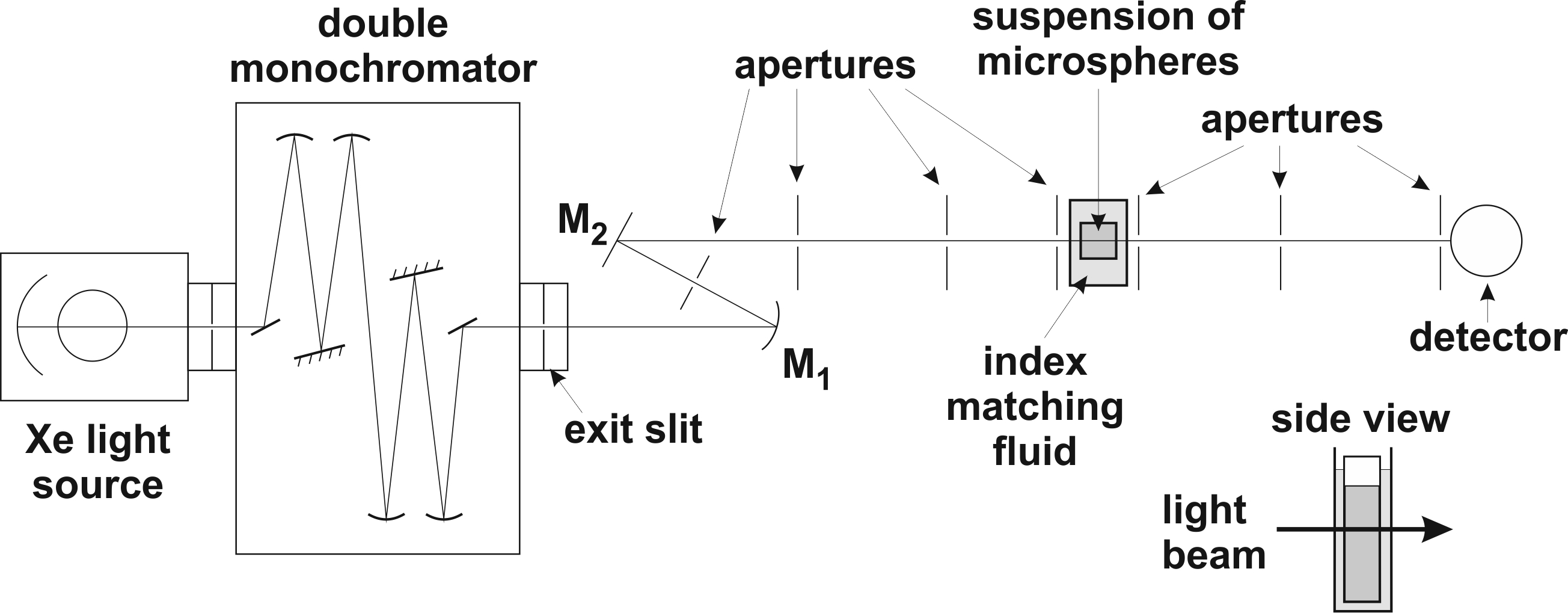}
 \caption{
  Schematic of experimental setup to measure spectral extinction cross sections of suspensions of spheres.
 }
 \label{fig:exp_setup}
\end{figure*}

\section{Introduction}

The refractive index (RI) describes the refraction of a beam of light at a (macroscopic) interface between any two materials.
Consequently, a variety of  experimental methods exist for measuring the RI of a material that
rely on the refraction or reflection of light at a planar interface between the
sample and some other known material, such as air, water or an optical glass.
This approach is feasible for materials that can exhibit macroscopically large defined interfaces, such as bulk liquids, homogeneous solids
or thin films and permits RI measurements with high accuracy.
For example, the refractive index of synthetic calcium fluoride has been determined with high accuracy between
\SI{138}{nm} and \SI{2326}{nm}  \cite{Daimon2002HighAccuracy} and for distilled water \cite{Daimon2007refractivewater} applying the minimum deviation method.
However, in many cases one is interested in the optical properties of materials that do not have a homogeneous macroscopic form, such as 
atmospheric aerosols \cite{Jones1994theoreticalstudy, Jones1994experimentalstudy, Thomas2005retrieval},
soot particles in flames \cite{Felske1992techniquedetermining}, biological cells or tissues \cite{naumenko1982determination, Ma2003polystyrenemicrospheres}
or man-made nanoparticles \cite{khlebtsov2008determination}. Often times, these materials cannot be condensed into  a homogeneous bulk sample.
If the size is comparable to the wavelength of visible light, media containing these particles (\eg, a suspension of
cells or a  cloud) are \emph{turbid}, \ie, light is scattered in  a complex process instead of propagating in straight rays.
Nevertheless, the interaction of light waves with such materials is still determined by the optical and geometrical properties of the constituting particles.
It is thus reasonable to ask the question ``What is the RI of the small particles contained in the inhomogeneous sample?''. This RI can yield information about the chemical composition of the particles
or  can be used for further mathematical modeling of light scattering processes. For similar purposes, one is interested in the size of the particles under consideration,
\eg, to calculate the force exerted by an optical trap \cite{Sun2008theory, Stilgoe2008effect},
For certain cases both properties can be inferred from measurements of the scattering and absorption of light by the particles.

A reference case is that of homogeneous spheres described by a single refractive index, since an analytical solution for the mathematical problem of
light scattering exists for this class of particles (Mie theory) \cite{mie1908beitrage, BohrenHuffman}. This makes the analysis of light scattering data feasible and at the same time is 
a good approximation for many real-world situations.
Thus far, several techniques to infer the RI or the size of spherical or small particles from measurements of the scattering or extinction of light have been discussed in the 
literature \cite{naumenko1982determination, Felske1992techniquedetermining, Jones1994theoreticalstudy, Jones1994experimentalstudy, Nefedov1997analysis, Ma2003polystyrenemicrospheres, Thomas2005retrieval, khlebtsov2008determination,
zhang2015influence, Blumel2016infrared}
These techniques can be divided into  those relying on the angular scattering pattern at a single wavelength \cite{naumenko1982determination, Jones1994theoreticalstudy, Jones1994experimentalstudy, Nefedov1997analysis},
and into those techniques relying on spectra of extinction or diffuse transmittance \cite{Ma2003polystyrenemicrospheres, Thomas2005retrieval, khlebtsov2008determination, zhang2015influence, Blumel2016infrared}.
Of course, a combination of these techniques is also possible \cite{Felske1992techniquedetermining}.

The inference problem or \emph{inverse problem} ``What is the RI and size distribution for a given extinction spectrum?''
is generally challenging.
Because the samples used and quantities measured for the various approaches described in the literature are so different,
researchers have come up with tailor-made solutions for their particular fields of application.
For example, in Ref.~\citenum{Ma2003polystyrenemicrospheres}, the authors used combined collimated and diffuse transmittance 
measurements to determine the RI of polystyrene  spheres of known size.
In Ref.~\citenum{Thomas2005retrieval}, the authors analyzed atmospheric aerosols of variable chemical composition. 
The complex refractive index was restricted to those functions represented by a multi-band damped-harmonic-oscillator model. 
This allowed to retrieve the RI of the aerosols. However, the  approach was unable to represent asymmetric dispersion features.
The method described in Ref.~\citenum{khlebtsov2008determination} was developed for nanoparticles and consequently could be applied only to particles of sub-wavelength size. Measurements in different suspending media were required for RI determination.
The authors of Ref.~\citenum{Blumel2016infrared} explicitly analyzed the location of ripples in the extinction spectrum of single microspheres using synchrotron radiation.
While the RI could be determined quantitatively from the the relative shape of these ripples, an accurate particle sizing was not possible.

In this work we present a method to infer the wavelength-dependence of the  real RI and the particle size distribution, characterized by mean value and standard deviation,
from a single measurement of spectral extinction cross section for spherical particles in the Mie scattering regime, \ie, the size is comparable to the wavelength.
We exemplify the method for suspensions of microscopic synthetic polystyrene (PS) beads that are widely used in colloidal and optical research, \eg, 
as a calibration material for cell measurements in optical and impedance flow cytometry \cite{Frankowski2015simultaneous}
or attached as ``handles'' to optically manipulate biological cells \cite{svoboda1994biological}. From the measurement of an extinction spectrum in the UV-VIS range with a tabletop experimental setup, 
one can infer the refractive index with an accuracy of 2 to 4 decimal places, depending in the wavelength, as well as the size parameters with accuracies in the range of one nanometer.

\section{Materials, methods and models}\label{sec:mat_meth_mod}
\subsection{Experiment}\label{sec:experiment}
\subsubsection{Experimental Setup}
To determine spectral extinction cross sections of polystyrene microspheres we measured the collimated transmission of particle suspensions using the setup depicted in Fig.~\ref{fig:exp_setup}.
As light source a \SI{1}{kW} Xenon high pressure discharge lamp is mounted to a double monochromator (model 1680 double spectrometer, SPEX industries, Inc., USA, NJ), equipped with a \SI{1200}{gr.mm^{-1}} grating 
for the wavelength range from \SI{185}{nm} to \SI{900}{nm}. The focal length of the double monochromator is \SI{22}{cm}. When measuring the collimated transmission of particle suspensions, the scanning range was reduced to \SI{260}{nm}--\SI{800}{nm}. 
Outside this wavelength region, the signal to noise ratio was insufficient, since below \SI{260}{nm} the output power of the Xe high pressure lamp is too low and above \SI{800}{nm} the sensitivity of the photomultiplier tube 
(model R928, Hamamatsu Photonics Deutschland GmbH, Germany) significantly drops. The slit width was set to obtain a spectral resolution of \SI{0.5}{nm} and the wavelength increment amounted to \SI{1}{nm}.
Spectra were recorded with an integration time of \SI{1}{s} and a scanning speed of \SI{0.5}{nm.s^{-1}}. Hence, the total measurement time was \SI{18}{min}.

The light emitted from the monochromator was collimated by a spherical mirror of \SI{20}{cm} focal length ($\mathrm{M_1}$) and reflected to the sample cell by a plane mirror ($\mathrm{M_2}$). 
The system of apertures served to minimize the divergence of the light beam and to reduce stray light and ambient light. The adjustable apertures were set to a diameter of \SI{2}{mm}.
Taking into account the distance of \SI{90}{cm} between the first aperture and the aperture directly mounted in front of the cuvette with the index matching fluid,
the maximum cone of light that can enter the cuvette is characterized by a half-angle of \SI{0.13}{\degree}. 
The suspension of microspheres was pipetted in a quartz glass cuvette (type 110-QS, Hellma GmbH \& Co. KG, Germany) used for photometric absorption measurements. 
The path length of the cuvette is $d=\SI{10}{mm}$ with an uncertainty of $u(d)=\SI{10}{\micro\meter}$. To avoid the influence of slightly different angles when inserting the cuvette in 
the beam path, which would result in changes in back reflection and transmission, a container filled with index matching oil (Immersol 518N, Carl Zeiss Microscopy GmbH, Germany) 
was permanently mounted. The size of the container (inner dimensions $\SI{20}{mm}\times\SI{30}{mm}$) was chosen in such a way that the quartz glass cuvettes, featuring outer dimensions of 
$\SI{12.5}{mm}\times\SI{12.5}{mm}$, could be easily changed. 
The long distance of \SI{120}{cm} between the cuvette and the detector, together with a system of adjustable apertures set to \SI{2}{mm} diameter, served to effectively suppress light scattered at small angles. 
The half-observation angle amounted to \SI{0.1}{\degree}. 
As stated above, the transmitted light was recorded using a R928 photomultiplier tube.

\subsubsection{Properties of polystyrene microspheres}
\paragraph{Diameter and size distribution}
To validate the experimental procedure and the associated mathematical model we measured the collimated transmission of two different types of PS microspheres.
The mean diameter of the first type (dynospheres\texttrademark{} 50.010.SS-021 P, LOT Q262, Dyno Particles A.S., Norway) -- the results are denominated as dataset 1 in the following --
was specified as \SI{2.0}{\micro m}. The stated coefficient of variation of 1.2\% corresponds to a standard deviation of \SI{0.024}{\micro m}. 
The mass of PS amounted to \SI{0.1}{g} in \SI{10}{mL}. Taking into account the density of PS of \SI{1.05}{g.mL^{-1}} and the particle volume of 
\SI{4.19}{fL} the concentration was $c = \SI{2.3e9}{mL^{-1}}$. 
As second sample (dataset 2) we chose microspheres (Flow Check\texttrademark{} microparticles, Cat. No. 23526-10, Polysciences Europe GmbH, Germany) 
with a slightly different diameter of \SI{2.076}{\micro m} and larger standard deviation of \SI{0.053}{\micro m} (CV = 2.6\%).
Since these particles are generally used for calibration purposes in flow cytometry without diluting the sample, 
the concentration of about $10^7\,\si{mL^{-1}}$ is significantly lower compared to the other sample used. 
\paragraph{Sedimentation}
The influence of sedimentation of the particles during the measurement time of \SI{18}{min} was estimated by calculating their velocity using Stokes’ law. 
In equilibrium, the frictional force and the excess force due to the difference in density of water and PS are equal. 
Using the density of PS $\rho_\text{PS}=\SI{1.05}{g.mL^{-1}}$ and the dynamic viscosity of water $\mu_{\rm H_2O} = \SI{9.321e-4}{kg.m^{-1}.s^{-1}}$ at \SI{23}{\celsius} the sinking velocity of
PS particles with \SI{2}{\micro m} diameter amounts to \SI{0.12}{\micro m.s^{-1}}.
It follows that the sedimentation path in \SI{18}{min} is about \SI{0.13}{mm}.
This value is negligible, since the light beam is positioned in the middle of the cuvette, the filling height of which is typically \SI{2}{cm}, and the depletion zone as well as the enrichment zone are in a distance of 
about \SI{1}{cm}.

\begin{figure}[tb]
 \centering
 \includegraphics[scale = 0.84]{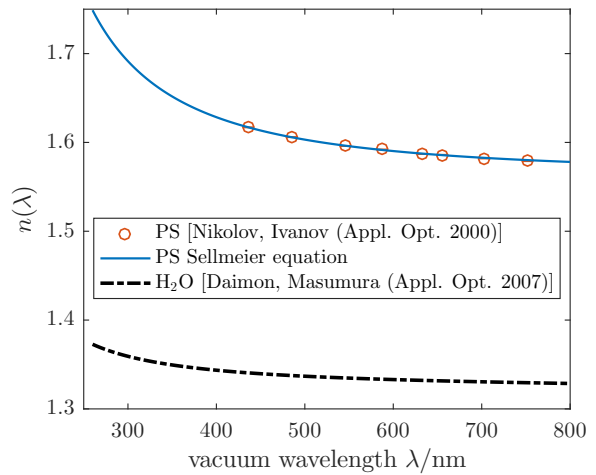}
 \caption{Refractive index of water \cite{Daimon2007refractivewater} and polystyrene \cite{Nikolov2000opticalplastic}. To create synthetic data, the data for PS was extrapolated from the measured range \SI{436}{nm}--\SI{1052}{nm} to the near UV
 by the Sellmeier equation shown as a blue solid line.}
 \label{fig:nH2O_nPS}
\end{figure}

\paragraph{RI of bulk polystyrene and water}
For the analysis of the experimental spectral extinction cross sections initial values are needed for the wavelength dependence of the refractive index of the PS microspheres. 
To this end, we employ  literature values for the RI of bulk PS and water, shown in Fig.~\ref{fig:nH2O_nPS}.
The data for the PS RI from Ref.~\citenum{Nikolov2000opticalplastic}  can be 
fitted very well using a one-term Sellmeier equation containing  a single absorption pole.%
\footnote{
  \label{footnote:Sellmeier}
  The one-term Sellmeier equation reads $n(\lambda)^2 = 1 + \frac{B\,\lambda^2}{\lambda^2 - C}$ and fits the 
  literature data for PS \cite{Nikolov2000opticalplastic}  with $B = 1.4432$ and $C = (\SI{142.1}{nm})^2$. }
We used this extrapolation curve to generate synthetic data for the spectral range [260, 800]\,\si{nm}.
For the wavelength-dependent RI of pure water we used a four-term Sellmeier equation \cite{Daimon2007refractivewater}
that is accurate to a few \num{e-6} in the wavelength range \SI{182}{nm}--\SI{1129}{nm}.
The absorbance of water in the wavelength range under consideration
is very low and can thus be neglected, hence $n_{\rm H_2O}\in \mathbb{R}$.
The same basically holds for PS, however, there is the onset of a (weak) UV absorption line at \SI{4.5}{eV} or \SI{275}{nm} \cite{Inagaki1977optical}.
Nevertheless, we treat its RI as real-valued.

\subsubsection{Sample preparation and transmittance measurements}
\begin{figure}[ht]
 \centering
 \includegraphics[scale = 0.84]{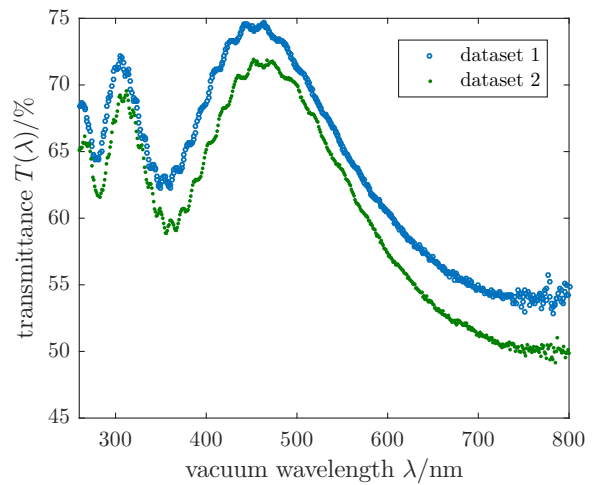}
 \caption{Transmission spectrum of polystyrene microspheres  with a diameter of approximately \SI{2}{\micro m}.}
 \label{fig:transmittance_exp_2um}
\end{figure}
The mathematical model for the analysis of extinction spectra requires that only single scatter events occur when the light passes the cuvette.
Multiple scattering is not included. To ensure that this condition is met, the concentration of particles is chosen correspondingly and measured 
by a flow cytometer, designed to determine reference values for concentrations of blood cells and particles in suspensions 
\cite{Ost1998eryleuko, Reitz2010determination}.
The mean free path length between two scattering events was estimated by taking two times the geometrical cross section as optical extinction cross section. 
For a transmittance of, \eg, 70\% the mean free path length of \SI{30}{mm} exceeds the cuvette path length by a factor of 3.
For the samples the dilution was adjusted to yield concentrations $c$ of $(4.95\pm0.09)\times{10^6}\,\si{mL^{-1}}$ and $(5.23\pm0.10)\times10^6\,\si{mL^{-1}}$ for datasets 1 and 2, respectively.
In Fig.~\ref{fig:transmittance_exp_2um}, we depict the result of  measurements of the spectral transmission for the two datasets. 
These spectra represent the transmittance of the particle suspensions, since blank values were corrected. 
To this end, we first measured the transmittance $I_0(\lambda)$ of the cuvette with pure water. Subsequently, the cuvette was cleaned and filled with the particle suspension. 
The transmittance $I(\lambda)$ of the suspension was measured and the spectral transmittance derived as ratio between both measurements:
\begin{equation}
 T(\lambda)= \frac{I(\lambda)}{I_0(\lambda)}.
\end{equation}
It should be noted that all measurements were normalized to the spectral output power of the monochromator using the quantum absorber \cite{melhuish1972absolute} integrated in the instrument.

The spectral transmittance can also be expressed as
\begin{equation}
 T = \exp\left(-\frac{C_\text{ext}^\text{ensemble}}{\Omega}\right),
\end{equation}
where $\Omega$ is the area illuminated by the beam.
$C_\text{ext}^\text{ensemble}$ is the extinction cross section of the particle ensemble.
The particles in the sample are
specified by the manufacturer as monodisperse,
since generally the variation in size is negligible for the intended purpose, \eg, calibration of flow cytometers. 
However, spectral extinction is sensitive to the size variation and consequently we have to consider the particles as
\emph{polydisperse}. 
Because we are dealing with incoherent single scattering events,
the total extinction cross section of the ensemble of $\nu$ particles is the sum of the single particles' extinction cross sections
\begin{equation}
 C_\text{ext}^\text{ensemble} = \sum_{j=1}^\nu C_\text{ext}^j.
\end{equation}
With the optical path length of $d=\SI{10}{mm}$, the particle concentration of $c\approx \SI{5e6}{mL^{-1}}$ and the area $\Omega = \SI{3.1}{mm^2}$ given by a \SI{2}{mm} aperture,
the number of particles $\nu = \Omega\,d\,c$ is on the order of $10^5$.
This  allows for the measurement of an ensemble average, denoted by
\begin{equation}
 \overline{C}_\text{ext} = \frac{1}{\nu} \, C_\text{ext}^\text{ensemble} = -\ln\left(\frac{I}{I_0}\right)\,\frac{1}{d\, c}. \label{eq:C_aver}
\end{equation}

The path length $d$ is known to a relative accuracy of $10^{-3}$ and the  particle concentration $c$ was measured using flow cytometry to a relative standard deviation of 2\%.
The latter  uncertainty contribution thus introduces an uncertainty into the  scaling factor for the measured ensemble average in Eq.~\eqref{eq:C_aver}.
In the following, we drop the subscript $_\text{ext}$, \ie, $\overline{C}_\text{ext}  = \overline{C}$ and denote experimental data by an asterisk: $\overline{C}^*$.

\subsubsection{Noise model for cross section measurements}\label{sec:exp_uncertainty}
\begin{figure}[t]
 \centering
 \includegraphics[scale = 0.84]{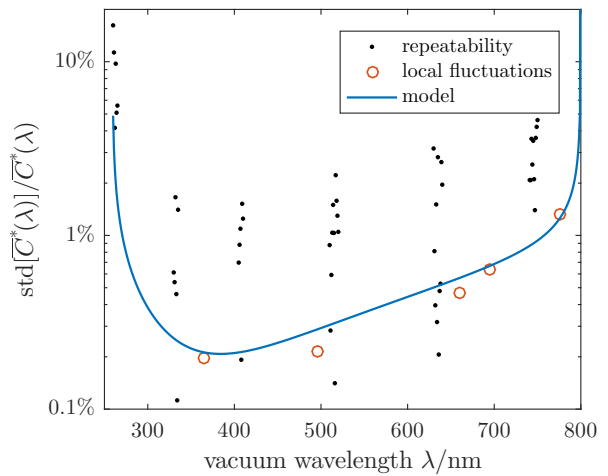}
 \caption{Estimates of the coefficient of variation $\hat{\sigma}[\overline{C}^*\!(\lambda)]$ 
 due to measurement noise: Maximal estimate from repeated measurements (black dots), minimal estimate
 from local fluctuations of $\overline{C}^*\!(\lambda)$ (red circles) and model curve [Eq.~\eqref{eq:bathtub_model}, blue line].
 }
 \label{fig:spectral_CV}
\end{figure}

There are different sources of measurement uncertainty:
(1) concentration uncertainties as discussed above, (2) detector noise, and (3) possible other systematic influences yet to be determined.
The latter point is discussed in subsection~\ref{sec:correction_syst_infl}.

We assume the detector noise to be white (\ie, uncorrelated between different wavelengths),
to have a wavelength-dependent amplitude $\std[\overline{C}^*\!(\lambda)]$, and a coefficient of variation 
\begin{equation}
 \hat{\sigma}[\overline{C}^*\!(\lambda)] = \frac{\std[\overline{C}^*\!(\lambda)]}{\overline{C}^*\!(\lambda)}.
\end{equation}
We estimated $\hat{\sigma}[\overline{C}^*\!(\lambda)]$ from the data in two different ways:
\begin{enumerate}
 \item From the variation of repeated measurements for the same sample in overlapping spectral regions, which yields an upper bound of the statistical measurement error,
 because it also contains systematic contributions resulting from drift of measurement conditions and other sources.
  The estimated $\hat{\sigma}[\overline{C}^*\!(\lambda)]$ has a characteristic ``bathtub shape'' (Fig.~\ref{fig:spectral_CV}), which we approximate by a function
  \begin{equation}
   \hat{\sigma}[\overline{C}^*\!(\lambda)] = \frac{p_1}{\sqrt{p_2^2 - (\lambda-p_3)^2} }+ p_4\,\lambda + p_5. \label{eq:bathtub_model}
  \end{equation}
  with five fit parameters. 
  Such a shape indicates increased noise contributions due to low output power of the light source in the UV and a decreasing sensitivity of the detector for long wavelengths.  
  \item  From the local fluctuations of $\overline{C}_\text{ext}$ in regions of vanishing slope, which yields an estimate of the
  measurement error. Due to the pronounced ripple structure of the data, no estimate is possible at the blue end of the 
  spectrum and the bathtub shape is not resolved. We use the same model function  [Eq.~\eqref{eq:bathtub_model}] as for the
  estimate based on the variation of repeated measurements above  and simply rescale it by a factor $<1$ to optimally fit the minimal error estimate. The resulting curve 
  is shown in Fig.~\ref{fig:spectral_CV} and is used for weighting and uncertainty analysis in the following.
\end{enumerate}

\subsection{Mathematical model}\label{sec:math_mod}
\subsubsection{Mie scattering}
The scattering particles are assumed to be spherical and multiple scattering effects are negligible due to the low particle concentration and low sample thickness.
Mathematically, the experiment can thus be described using the time-harmonic Maxwell equations for the scattering
of a plane electromagnetic wave by a single homogeneous sphere. This problem has an analytic solution, which is  known as \emph{Mie theory} or \emph{Mie scattering} \cite{mie1908beitrage, BohrenHuffman}.
Mie scattering yields the full electric field for the scattering problem as a series expansion in vector spherical harmonics, from which, among other quantities, the total extinction cross section $C_\text{ext}$ can be computed.
The problem is fully characterized by two parameters:
\begin{enumerate}
 \item The size parameter \[
                            X := \frac{2\pi\, \nmed}{\lambda}\, R,
                           \]
  where $R$ is the radius of the sphere, $\lambda$ is the wavelength in vacuo 
  and $\nmed\in\mathbb{R}$ is the RI of the (non-absorbing) medium (water), respectively.
 \item The relative refractive index
 \[
  \mathfrak{m} := \frac{\mathfrak{n}}{\nmed} \in \mathbb{C},
 \]
 where $\mathfrak{n} = n+\imag \kappa\in\mathbb{C}$ is the RI of the sphere. The imaginary part $\kappa$  describes the absorption of light in the sphere's material. 
\end{enumerate}

Consequently, besides the known quantities vacuum wavelength $\lambda$ and RI of the medium $\nmed(\lambda)$,
the two quantities sphere radius $R$ and sphere's RI  $\mathfrak{n}(\lambda)$ are the two free parameters of the system.

Using a numerical implementation of the Mie scattering formulae \cite{WiscombeMieCode}, accurate values for the extinction cross section of a single particle are easily obtained for any given parameters $\mathfrak{n}(\lambda)$ and $R$
in the relevant range.
We denote these  numerically obtained values by the function
\[
 \mathcal{C}(\lambda; \mathfrak{n}, R).
\]
This notation shall indicate that we are considering  a wavelength-dependent function with two parameters, one of which, the RI $\mathfrak{n}(\lambda)$, is a function of the wavelength itself.

\subsubsection{Polydisperse ensembles}
Since the number of particles measured simultaneously in the experiment is large, the ensemble average in Eq.~\eqref{eq:C_aver} is
modeled by  simply integrating over the corresponding size distribution 
\begin{align}
 \overline{\mathcal{C}}(\lambda; \mathfrak{n}|r) = \int_{0}^\infty \mathcal{C}(\lambda; \mathfrak{n}, R)\, r(R)\,\dif R,
\end{align}
where $r(R)$ is the probability density function (pdf) of the radius $R$. 
We model it by a normal distribution
\begin{equation}
  r(R) \propto \exp\left\{ -\frac12 \frac{[R/R_c-\mu_R]^2}{\sigma_R^2} \right\}.
\end{equation}
Here $R_c$ is a typical length scale (\eg, a rough guess for the mean particle radius),
rescaling the parameters to dimensionless quantities of the order of 1.
$\mu_R$ is the mean and $\sigma_R$ the standard deviation of the distribution of radii, relative to the characteristic scale $R_c$.
We combine these two parameters of the distribution  into the vector 
\begin{equation}
 \bm{\theta} := (\mu_R,\sigma_R)^T.
\end{equation}

A normal distribution  was chosen as a model function since it is known to describe the size distribution of polystyrene microspheres very well \cite{lettieri1991certification}.
Alternatively, we also tested a log-normal distribution. For the narrow size distributions of the samples used here, the difference in shape between a normal distribution
an  a log-normal distribution
of identical mean value and standard deviation is only minor. Consequently, the results were virtually the same in both cases.

To implement the integrals in the ensemble average numerically, we 
used trapezoidal sums over uniform grids, with appropriate spacing and range.

\subsubsection{Correction for systematic influences}\label{sec:correction_syst_infl}
Including concentration errors in the forward model for the measured curve would result in the ensemble averaged $\overline{\mathcal{C}}(\lambda; \mathfrak{n} | r)$ being multiplied with a
prefactor $1-\epsilon_c$, where $\epsilon_c$ corresponds to the relative concentration error with $\mathbb{E}(\epsilon_c) = 0$, $\std(\epsilon_c) \approx 2\%$.
However, it turns out that the forward model is much more flexible and the quality of the solution of the inverse problem can be greatly enhanced when
a simple parametric wavelength dependence  of this prefactor is introduced, \ie, instead of $1-\epsilon_c$, we have $1-\epsilon_c + \epsilon_o(\lambda; \bm{\eta})$,
where $\epsilon_o(\lambda; \bm{\eta})$ corrects for influences \emph{other} than the concentration.
Mathematically, the influences of concentration and other factors can not be distinguished, hence we merge them into a single wavelength-dependent function $\epsilon(\lambda; \bm{\eta}) = -\epsilon_c + \epsilon_o(\lambda; \bm{\eta})$.
This \emph{compensation curve} is characterized by the parameters $\bm{\eta} = (\eta_1,\dotsc,\eta_{m_\eta})^T$
and defined as a linear interpolation between a few grid points
\begin{equation}
 \epsilon(\lambda; \bm{\eta}) := \begin{cases}
                          \eta_j & \lambda = l_j\\
                          \text{linearly interpolated } & \text{between the }l_j\\
                          \eta_1 & \lambda<l_1\\
                          \eta_{M_\eta} & \lambda>l_{M_\eta}
                         \end{cases},
\end{equation}
with $l_1<\dots<l_{M_\eta}$, $\bm{\eta} = (\eta_1,\dotsc,\eta_{M_\eta})^T$. 
A small number of points ($M_\eta = 4$ in our case) is  sufficient to significantly improve the
quality of the fit compared to the one obtained with a constant function.
Hence, our model for the measured $\overline{C}^*\!(\lambda_j)$ is
\begin{equation}
  \mathcal{M}(\lambda; \mathfrak{n}, \bm{\theta}, \bm{\eta}):=[1+\epsilon(\lambda; \bm{\eta})]\,\overline{\mathcal{C}}(\lambda; \mathfrak{n} | r).
\end{equation}
From a physical point of view, this wavelength-dependent factor serves to correct  for one or several influences, which are difficult to quantify here,
in particular aspects such as non-sphericity as well as the finite-size detector aperture and the beam divergence.
From a mathematical point of view, the introduction of $\epsilon(\lambda; \bm{\eta})$ is justified by the much higher quality of
the fit. As will be discussed in section~\ref{sec:res}, we obtain more convincing results for the fitted RI and size distribution
using this correction. 

For a choice of the grid points $l_1,\dotsc,l_{M_\eta}$ we tested uniform grids as well as a random placement using $M_\eta=0..10$ points. However, the best results 
(measured by the quality of the fit) were obtained when the local minima and maxima
of the slow oscillations of $\overline{C}^*\!(\lambda)$ were selected, which are at  \SI{300}{nm}, \SI{350}{nm},  \SI{450}{nm} and \SI{770}{nm} for \SI{2}{\micro m} 
PS spheres in water (Fig.~\ref{fig:Cext_exp_2um}).

\subsection{Inverse problem}\label{sec:inv_prob}
We now have a mathematical model for the measurement and are able to compute the average extinction cross section  for a given ensemble
of spheres, \ie, for a given size distribution and RI $n(\lambda)$.
We assume $\mathfrak{n}(\lambda) =  \Re\left[\mathfrak{n}(\lambda)\right] =  n(\lambda)$ here and in the following
due to  polystyrene's negligible absorbance.

The question we now address is: Can one infer the RI $n(\lambda)$ from measurements $\left[\overline{C}^*\!(\lambda_j)\right]_{j=1}^N$? First of all, it is clear that 
one should not try to infer more than $N$ parameters from $N$ given measurement data. This leaves the possibility to obtain $n(\lambda_j)$ for all $\lambda_j, j=1..N$, given 
knowledge about the size distribution
and particle concentration,
which is  a relatively simple problem: At each wavelength $\lambda_j$,
find $n_j = n(\lambda_j)$, such that 
\begin{equation}
 \overline{\mathcal{C}}(\lambda_j; n_j | r) \stackrel{!}{=} \overline{C}^*\!(\lambda_j). \label{eq:pointwise_approach_nec_cond}
\end{equation}
The problem is hence stated as finding roots of a nonlinear equation for all wavelengths separately. Hence, this approach is
\emph{pointwise}.
However, as it turns out, the pointwise approach fails: Slight, sub-percent errors in the size distribution or  in the particle concentration strongly affect the result of the RI reconstruction
and Eq.~\eqref{eq:pointwise_approach_nec_cond} may not even have solutions any more for some wavelengths. 
Since the prior knowledge about the parameters of the size distribution is not accurate enough, they
 need to be inferred from the data as well.
At first glance, this leaves one with more parameters to be reconstructed than data points. But the problem can be restated in a non-pointwise sense as a least-squares optimization problem.

\subsubsection{Representation of $n(\lambda)$}
We restate the problem by implying a constraint on the admissible functions $n$, namely that they can be expressed as
a finite  sum of smooth basis functions $g_j$, \ie,
\begin{align}
 n(\lambda) = \sum_{j=1}^M a_j g_j(\lambda)\quad\forall\,\lambda\in\mathbb{R} \label{eq:RI_series_expansion}
\end{align}
with unknown coefficients $a_j$.
Hence, we are working in the subspace spanned by the $M$ basis functions with  $M<N$.

Simple oscillator models for light-matter interactions reveal that particles with a resonance at wavelength
$\Lambda>0$ and of width $\gamma>0$ exhibit an \emph{anomalous dispersion} feature described by the expression
\begin{equation}
 n(\lambda)-1 \propto \frac{\lambda-\Lambda}{(\lambda -\Lambda)^2+\gamma^2},
\end{equation}
which is thus a generic shape of a feature of the real part of the RI.
Hence, we represent the refractive index $n(\lambda)$ in a sparse form by working in the $M$-dimensional space
\begin{equation}
 n \in \mathrm{span}\{1, f_1,\dotsc,f_{M-1}\},
\end{equation}
where 
\begin{equation}
 f_j(\lambda) :=\frac{\lambda-y_j}{(\lambda-y_j)^2+\gamma^2}.\label{eq:basis_nonb}
\end{equation}
The centers of the peaks $y_j$ are uniformly spaced, \ie, $y_{j}-y_{j-1} = \Delta_y$. 
We chose a grid spacing of $\Delta_y = \SI{30}{nm}$ and a  constant width of $\gamma = \SI{100}{nm}$ for all functions. The first (last) grid point 
was set to be one grid spacing $\Delta_y$ smaller (larger) than the lowest (highest) wavelength.

These functions $f_i$ are linearly independent, but they  are not useful for a practical implementation, since
they are far from being orthogonal, \ie, the scalar products between them are not close to zero, which would lead to problems in the numerics.
To avoid this problem, we orthonormalize the set of functions $\{1, f_1,\dotsc,f_{M-1}\}$, or rather the set the values of these functions at $(\lambda_1, \dotsc,\lambda_N)$, 
using the (modified) Gram-Schmidt process,
yielding an orthonormal set
$\{g_1,\dotsc,g_M\}$, \ie,
\begin{equation}
 \langle g_i, g_j\rangle = \delta_{ij}\text{\quad{}and\quad} g_1(\lambda_j) = \frac{1}{\sqrt{N}}\;\forall\,j=1, .., N,
\end{equation}
where $\langle .\,, .\rangle$ denotes the standard inner product in $\mathbb{R}^N$.

This representation of $n(\lambda_j)$ as a series expansion in adequately-chosen basis vectors
results in a reduction of the dimensionality by a factor of 30: 
Instead of one coefficient every \SI{1}{nm} corresponding to the spectral resolution,
only one coefficient every \SI{30}{nm} is needed.
At the same time, it is possible to represent the RI  of bulk PS (Fig.~\ref{fig:nH2O_nPS}) to machine precision.
But also curves less generic than this Sellmeier equation (see footnote 
\footnotemark[\value{footnote}]), \eg, the feature-rich spectral RI of the protein complex hemoglobin 
\cite{friebel2006modelfunction, gienger2016determining}
can be represented with errors well below the respective measurement uncertainties using an appropriate grid spacing $\Delta_y$.

\subsubsection{Nonlinear least-squares optimization}
\label{sec:LSQ}
\begin{figure}[t]
 \centering
 \includegraphics[width = \linewidth]{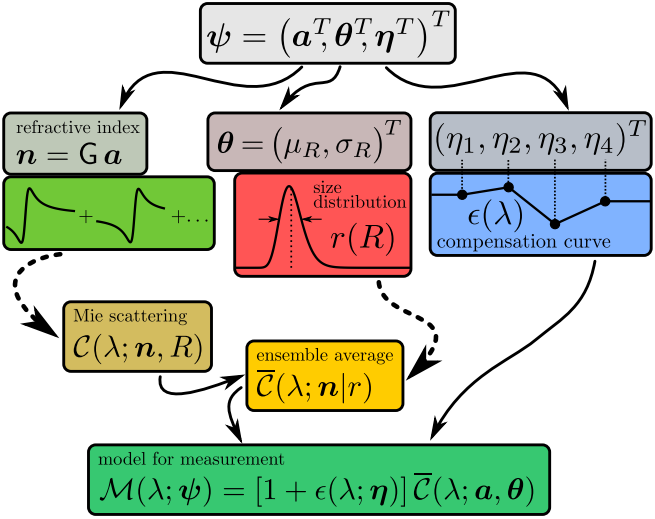}
 \caption{Schematic representation of the mathematical model of the experiment. Linear transformations are shown as solid lines, nonlinear dependencies as dashed lines.
 }
 \label{fig:model_schematic}
\end{figure}
The series expansion of the real RI in Eq.~\eqref{eq:RI_series_expansion}
can also be written as a  matrix-vector product
\begin{equation}
 \bm{n} = \mathsf{G}\,\bm{a}
\end{equation}
with $\mathsf{G} := \{g_j(\lambda_i) \}_{i,j}\in\mathbb{R}^{N\times M}$, $\bm{n} = (n(\lambda_1), \dotsc,n(\lambda_N))^T $.
The coefficient vector $\bm{a}$ describing the RI, the vector $\bm{\theta}$ describing the size distribution of the spheres
and vector $\bm\eta$ describing the compensation curve
are the unknown quantities which we need to recover from the spectral extinction data. Thus we combine them into a single parameter vector
\begin{equation}
 \bm{\psi}: = \begin{pmatrix}
               \bm{a}\\ \bm{\theta} \\ \bm{\eta}
              \end{pmatrix}
              \in\mathbb{R}^L,
\end{equation}
where $L=27$ with the parameters chosen for measurement data discussed here ($N=540$ points).
The mathematical forward model to compute  the extinction cross sections $\mathcal{M}(\lambda; \bm{\psi})$  from the parameter vector $\bm{\psi}$ is depicted in Fig.~\ref{fig:model_schematic}.

We then set up a quadratic cost function consisting of the summed squared residuals
\begin{align}
 F_i &:= \mathcal{M}(\lambda_i; \bm{\psi}) - \overline{C}^*(\lambda_i) \\
 \chi^2(\bm{\psi}) &:= \sum_{i,j=1}^N w_{ij}  F_i F_j = \bm{F}^T\,\mathsf{W}\,\bm{F},
\end{align}
where $\mathsf{W}  = \mathsf{V}(\bm{\overline{C}^*})^{-1}$ is a symmetric weight matrix
with
\begin{equation}
 \mathsf{V}(\bm{\overline{C}^*}) = \diag(\sigma_1^2,\dotsc, \sigma_N^2).
\end{equation}
The standard deviations $\sigma_i = \std[\overline{C}^*\!(\lambda_i)]$ are estimated according to Eq.~\eqref{eq:bathtub_model}
for each dataset.
The necessary condition for optimal parameters that minimize $\chi^2$ is then
\begin{equation}
 0\stackrel{!}{=}  \nabla_{\bm\psi} \chi^2(\bm{\psi}) = \mathsf{J}^T\nabla_{\bm F} \chi^2(\bm{\psi}) = 2\,\mathsf{J}^T \mathsf{W} \bm{F},
 \label{eq:optim_necc_cond}
\end{equation}
where we have introduced  the Jacobian matrix
\begin{equation}
 \mathsf{J} = \left\{  \frac{\partial F_i}{\partial \psi_j}  \right\}_{i j} \in\mathbb{R}^{N\times L}.
\end{equation}

Further details for the implementation of the forward model are given in Appendix~\ref{app:expressions}.
Numerically, the necessary condition can be solved by iterative local optimization methods. We tested both the Levenberg-Marquardt algorithm
and the trust-region-reflective algorithm as they are implemented in Matlab (MATLAB R2016a, The MathWorks, Inc.).
Both algorithms basically yield the same result when converged. Since the trust-region-reflective algorithm
proved  more stable with respect to the choice of initial values, all the results presented here were obtained 
with this algorithm.
When converged, the numerical routine yield an optimal parameter set $\hat{\bm{\psi}}$ corresponding to a local minimum and
$\chi^2$ normalized to the degrees of freedom
\begin{equation}
 \chi^2_{\rm dof}:= \frac{\chi^2(\hat{\bm{\psi}})}{N-L+1}
\end{equation}
provides  a measure for the quality of the fit.

\paragraph{Influence of initial parameter values}
The local nonlinear optimization employed for the same dataset can result in different estimates of the 
parameter vector $\bm{\psi}$ when different initial values are used. 
Firstly, it is possible that the found local minimum is far away from true parameter values (for synthetic data) or plausible parameter values (for experimental data).
This is usually indicated by a large value of $\chi^2_{\rm dof}$ (\ie, $\chi^2_{\rm dof}\gg1$). Secondly, one can also find different local minima
rather close to each other and -- in the case of synthetic data --
similarly close to the correct value, indicating multiple minima of the least-squares problem. 
For these minima further iteration does not improve the fit.
We estimated the resulting uncertainty stemming from this  effect  as follows: Different initial
parameter values were created  by adding normally distributed random numbers to the same mean initial parameter vector.
We tuned the amplitude of these random numbers such that the range of initial conditions is sufficiently wide and that, on the other hand, the 
optimization converged for the  majority of samples. However, very high values of $\chi^2_{\rm dof}$ do occur in some cases.
In order to include all of these samples appropriately, we introduced weights proportional to $\exp(-\chi^2_{\rm dof})$ in 
the averages and covariances over the ensemble of optimization runs. This penalty term attenuates samples with poor agreement between model and data
(\eg, $\chi^2_{\rm dof}=10$).
The result for the parameter vector is then obtained as the weighted average  $\langle\bm{\psi}\rangle_\text{init}$ over the ensemble of random initial parameter values.

The uncertainty analysis for the estimated parameters is presented in Appendix~\ref{app:unc_analysis}.

\section{Results}\label{sec:res}
\begin{figure}[ht]
 \centering
 \includegraphics[scale = 0.84]{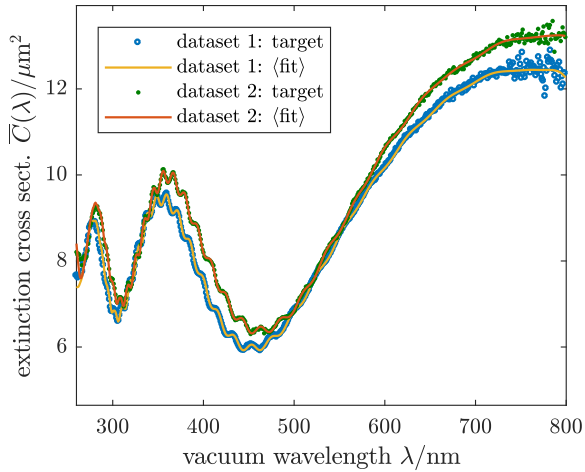}
 \caption{Experimental extinction cross sections (``target'') and model fits with optimized  parameters as described in subsections \ref{sec:math_mod} and \ref{sec:inv_prob}.
 See Data Files 1 and 2 for underlying values of the respective datasets.
 }
 \label{fig:Cext_exp_2um}
\end{figure}

\subsection{Inverse problem results}
To evaluate the method, we created synthetic datasets for polystyrene particles suspended in pure water. 
These synthetic data were created as follows: The optical properties 
assumed are shown in Fig.~\ref{fig:nH2O_nPS}.
The compensation curve  was fixed to $\epsilon(\lambda; \bm{\eta})\equiv 0$ in the forward model and Gaussian white noise was added according to the coefficient of 
variation estimated in Eq.~\eqref{eq:bathtub_model} and shown in Fig.~\ref{fig:spectral_CV}.
We analyzed synthetic data with $\mean(R)  = \SI{1}{\micro\metre}$, $\std(R) = \SI{5}{nm}$
for  different realizations of measurement noise and observed that the deviations of the found size distribution parameters were within the margins expected from
the corresponding estimated uncertainties. 
We computed the difference between the  RI values  $n(\lambda_j)$ obtained by optimization and those according to the  target curve.
The percentages of values within one (two) standard deviation(s) estimated from the uncertainty analysis were found to be close to 
the  68\% (95\%) one expects from a normal distribution, indicating a reasonable estimate of the uncertainties.
Generally, these percentages were even higher, indicating that the uncertainties are slightly overestimated.
The compensation curve also returned the target ($\bm{\eta} = 0$) within the uncertainties.

\begin{figure}
 \centering
  \includegraphics[scale = 0.84,
  clip=true,
  trim = 0 5.4mm 0 0
  ]{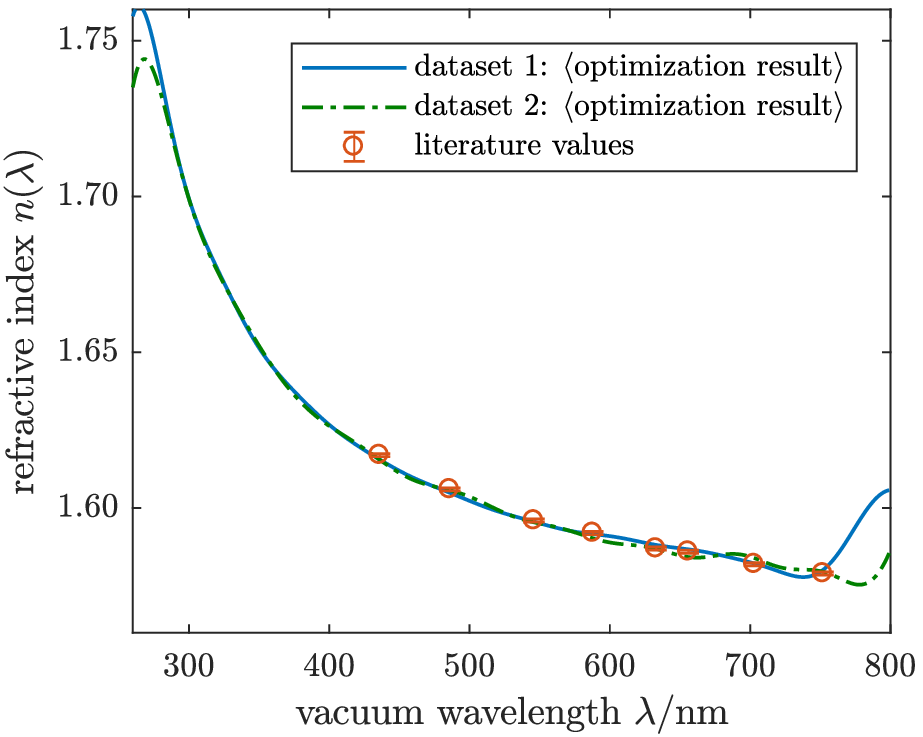}\\
  \textbf{(a)}
  \label{fig:nPS_exp_2um}
 \\[12pt]
    \includegraphics[scale = 0.84,
      clip=true,
      trim = 0 5.4mm 0 0
    ]{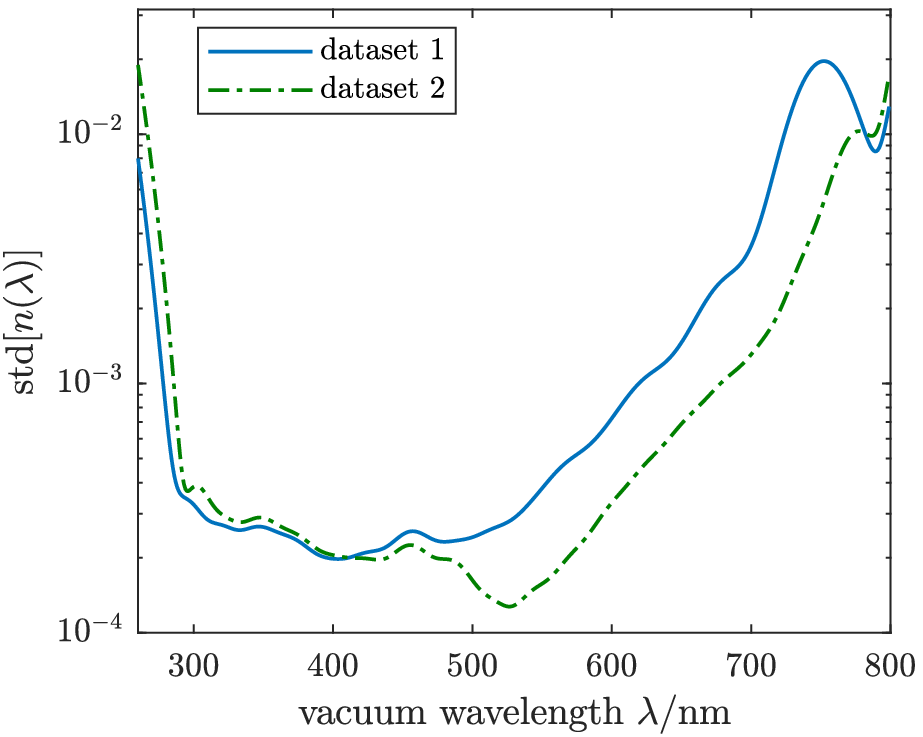}\\
  \textbf{(b)}
  \label{fig:std_nPS_exp_2um}
 \\[12pt]
    \includegraphics[scale = 0.84]{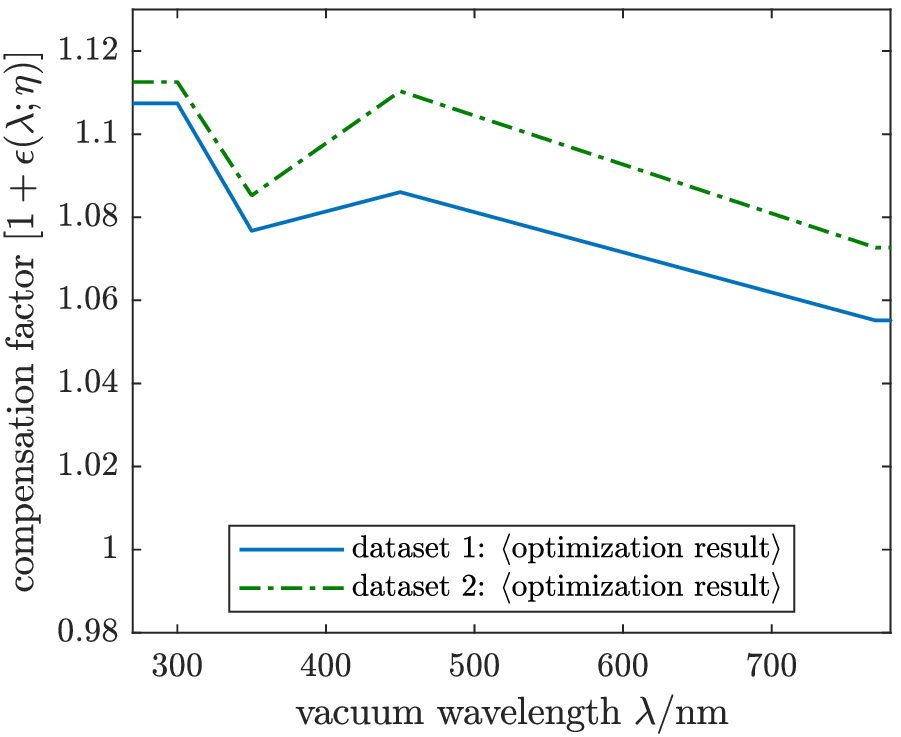}\\
   \textbf{(c)}
    \label{fig:u_exp_2um}
 
 \caption{Optimization results for the model parameters for \SI{2}{\micro \metre} PS spheres for two experimental datasets:
  \textbf{(a)}
  Refractive index.
  Literature values are from Ref.~\citenum{Nikolov2000opticalplastic}.
  The oscillations of $n(\lambda)$  at the red and UV end of the spectrum are within the estimated uncertainties (Fig.~\ref{fig:results_exp_2um}\,(b)).
   See Data Files 1 and 2 for underlying values of the respective datasets.
  \textbf{(b)}
   Estimated uncertainties (one standard deviation) of the RI (Fig.~\ref{fig:results_exp_2um}\,(a)).
  \textbf{(c)}
   Compensation curve. The piecewise linear function $\epsilon(\lambda;\bm{\eta})$ is spanned by 4 grid points at \SI{300}{nm}, \SI{350}{nm}, \SI{450}{nm} and \SI{770}{nm}.
 }
 \label{fig:results_exp_2um}
\end{figure}

The experimental cross sections, \ie, datasets 1 and 2 (Fig.~\ref{fig:Cext_exp_2um}), were analyzed in the same way.
The resulting RI is depicted in Fig.~\ref{fig:results_exp_2um}\,(a) and the corresponding uncertainty (Fig.~\ref{fig:results_exp_2um}\,(b)) is shown to be less than 1\% almost everywhere
and to be less than 0.1\% between \SI{300}{nm} and \SI{650}{nm}.
The compensation curve is shown in Fig.~\ref{fig:results_exp_2um}\,(c). The wavelength-dependence of $\epsilon(\lambda;\bm{\eta})$
is significant with the estimated uncertainties.
Tab.~\ref{tab:size_exp_2um} lists the mean values and standard deviations of the particle size distributions.
All these results correspond to the MC-averaged optimized parameters.

\begin{table}[b]
  \caption{Optimization results for size distribution parameters, cf. Fig.~\ref{fig:results_exp_2um}. 
  The numbers in parentheses are the standard uncertainties referred to the last digits of the respective results.
  }
  \label{tab:size_exp_2um}
  \centering
    \begin{tabular}{| l | S[table-format=4.4] |  S[table-format=4.4]|}\hline
    		&  \multicolumn{1}{c}{$\mean(R)/\si{nm}$} 	& \multicolumn{1}{|c|}{$\std(R)/\si{nm}$}\\\hline
    \multicolumn{3}{|c|}{dataset 1}\\\hline
     result 	&  1015.1(8)  	& 4.7(2)\\\hline
     specification& 1000 	&  12\\\hline      
    \multicolumn{3}{|c|}{dataset 2}\\\hline
     result 	&  1038.5(9)  	& 4.4(2)\\\hline
     specification&  1038 	&  27\\\hline      
    \end{tabular}
\end{table}

For wavelengths larger than \SI{436}{nm}, we compared the RI uncertainties with the deviations between the found RI curves and the Sellmeier curve for the literature RI of bulk PS 
(see footnote 
\footnotemark[\value{footnote}]). 
For the first dataset 
52\% (65\%) 
of the values were within one (two) estimated standard deviations of the literature values. For the second dataset
37\% (57\%) 
were within one (two) standard deviations.
Although these percentages are somewhat too low, it indicates that the estimated uncertainties are not too far off.
Furthermore, these numbers do not take into account any uncertainties of the RI literature values or possible deviations of the RI of PS microspheres in our samples from that of bulk PS.

As an additional consistency test, we computed the difference between the RI curves obtained independently from the two experimental datasets. 
This difference was within the combined estimated uncertainties, thus indicating that the 
uncertainty estimates are reliable.

\subsection{Effect of the compensation curve}

\begin{figure}[thb!]
 \centering
    \includegraphics[scale = 0.84,
     clip = true,
     trim = 0 5.4mm 0 0
    ]{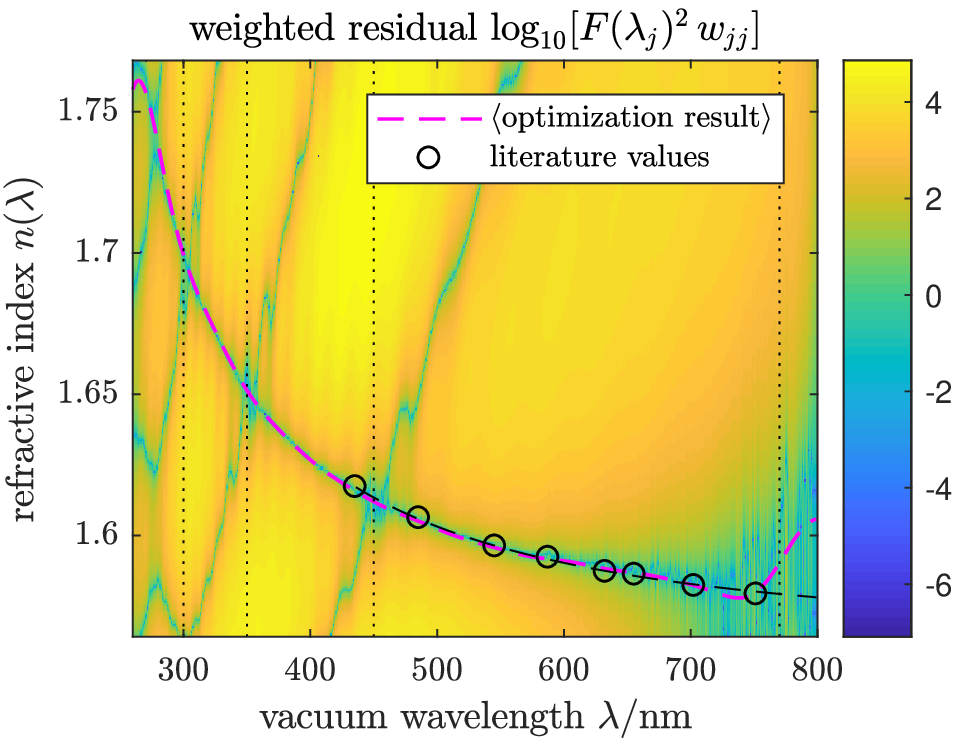}\\
    \textbf{(a)}
    \label{fig:local_res_nPS_exp_2um}
 \\[12pt]
   \includegraphics[scale = 0.84,
     clip = true,
     trim = 0 0 0 5mm
   ]{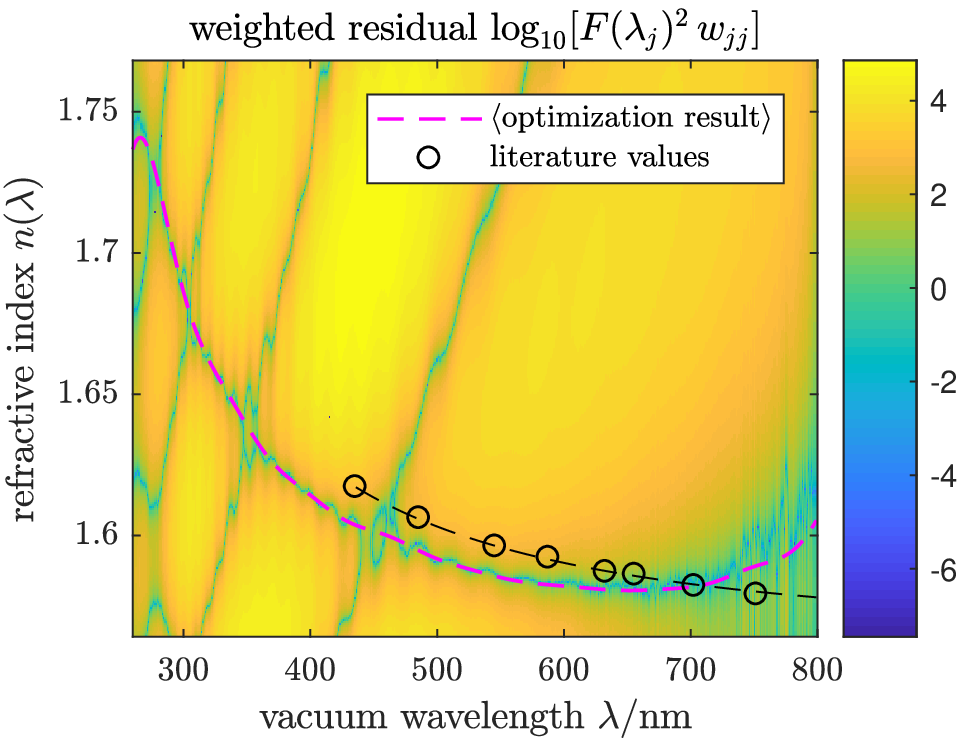}\\
 \textbf{(b)}
 \label{fig:local_res_nou_nPS_exp_2um}
 \caption{Dataset 1: Residuals $F(\lambda_j; \bm{\psi}) = \mathcal{M}(\lambda_j; \bm{\psi}) - \overline{C}^*(\lambda_j)$
    of the extinction cross section as a function of wavelength and assumed refractive index.
    The parameters $\bm{\theta}$ and $\bm{\eta}$ are fixed to the optimization result.
  \textbf{(a)}
    With forward model incorporating the compensation curve $\epsilon(\lambda;\bm{\eta})$ (Fig.~\ref{fig:model_schematic}). 
    The black dotted lines denote the wavelength grid spanning $\epsilon(\lambda;\bm{\eta})$ (Fig.~\ref{fig:results_exp_2um}\,(c)). Note the false solutions intersecting at the local extrema of 
    $\overline{C}^*\!(\lambda)$ (Fig.~\ref{fig:Cext_exp_2um}) and the blurred valley at the red end due to measurement noise.
  \textbf{(b)}
    With forward model not incorporating the compensation curve $\epsilon(\lambda;\bm{\eta})$.
    }
    \label{fig:local_res_nPS}
\end{figure}

In Fig.~\ref{fig:local_res_nPS}\,(a) we plot the local extinction-cross-section residuals $F(\lambda_j; \bm{\psi}) = \mathcal{M}(\lambda_j; \bm{\psi}) - \overline{C}^*(\lambda_j)$
as a function of wavelength and refractive index.
We performed an additional optimization with a forward model not including the compensation curve, \ie, $\bm{\eta}\equiv0$ in the optimization.
The result is shown in Fig.~\ref{fig:local_res_nPS}\,(b).
Comparing Figs.~\ref{fig:local_res_nPS}\,(a) and (b),
ones sees how the incorporation of $\epsilon(\lambda;\bm{\eta})$ into the model enables it to close the gaps that are otherwise present near the 
local extrema of  $\overline{C}^*\!(\lambda)$ (compare Fig.~\ref{fig:Cext_exp_2um}).
The ripples in  $\overline{C}^*\!(\lambda)$ are not matched by the model fit (data not shown) and for dataset 1 we obtained 
$\langle\chi^2_{\rm dof}\rangle= 7.3$ 
(indicating a poor fit) instead of
$\langle\chi^2_{\rm dof}\rangle= 0.78$ (indicating a very good fit) with the forward model including  $\epsilon(\lambda;\bm{\eta})$.
For dataset 2 these values were 
$\langle\chi^2_{\rm dof}\rangle= 10.6$ 
and 
$\langle\chi^2_{\rm dof}\rangle= 1.57$, 
respectively.
Furthermore, the curves for $n(\lambda)$ obtained for datasets 1 and 2 do not match
each other within the estimated uncertainties in the former case and neither of them matches the literature data.
In conclusion, the optimization with the forward model including $\epsilon(\lambda;\bm{\eta})$ (\ie, in our example with $L=27$ free parameters instead of $L=23$) yield much more consistent 
and plausible results than with the one not including $\epsilon(\lambda;\bm{\eta})$ and thus was chosen even though the physical origin of the effect cannot be identified unambiguously.

\subsection{Reliability of solutions for wider ensembles}
The procedure for solving the inverse problem and related uncertainty analysis works well for
the cases discussed here with $\mean(R)  \approx \SI{1}{\micro\metre}$ and $\std(R)/\mean(R)\approx 0.5\%$, \ie, for a narrow size distribution.
Using synthetic data, however, we found that the parameter retrieval becomes increasingly difficult for wider size distributions. 
This is because the characteristic ripples in $\overline{C}(\lambda)$ get smoothed out for wider size distributions.
This ripple structure is very sensitive to particle size and RI \cite{Blumel2016infrared}.
Hence, when this characteristic fine structure is suppressed by ensemble averaging essential information is lost. The effects of
RI, mean radius and particle concentration in the model can then compensate each other, which can lead to a good fit of $\overline{C}(\lambda)$ with systematically incorrect parameters.
To quantify this, we analyzed synthetic data with $\mean(R)  = \SI{1}{\micro\metre}$ and distribution widths of  $ 0.1\%\le \std(R)/\mean(R)  \le 5.0\%$.
Without added measurement noise, the correct parameters can  be found in all cases, \ie, 
this is not a problem of (numerical) sensitivity in the forward model.
In the presence of measurement noise, however, systematic parameter errors occur for values of $\std(R)/\mean(R) \ge 1\%$, because in these cases the sensitivity is obscured by measurement noise. 
These deviations typically increase with distribution width.
A similar effect can also be expected for low-contrast particles, because this attenuates the ripple structure as well.
This problem could in principle be resolved by achieving a sufficiently low measurement noise.

\section{Summary and discussion}\label{sec:summ_disc}
 We presented measurements of the spectral extinction cross sections of polystyrene microspheres suspended in water. Theses measurements
$\overline{C}^*\!(\lambda)$
represent averages over the size distribution of the polydisperse ensemble. Using a numerical implementation of Mie scattering, in combination with an appropriate description of the size distribution
and the spectral refractive index $n(\lambda)$ in a low-dimensional set of orthonormal basis functions, we developed a forward model $\mathcal{M}(\lambda;\bm{\psi})$ to mathematically describe the measurements
for a given parameter set $\bm{\psi}$. Applying standard nonlinear least-squares optimization, the $L$ model parameters contained in the vector $\bm{\psi}$ were fitted to $N$ measurement data points.
In our example, we had $N=540$ and $L=27$.
This yields the
spectral refractive index of the microspheres,  the mean value of their size and the  standard deviation of their size distribution.
The uncertainties of these results were estimated
using linearized propagation of covariance matrices and Monte Carlo sampling.

The wavelength dependence of the RI of PS microspheres derived by modeling experimental extinction cross sections including uncertainties 
yields results consistent with literature values. This good accordance proves that the microspheres are produced with a homogeneity in density and optical properties that is comparable to bulk material.
On the other hand, both samples feature about the same size distribution with a coefficient of variation of about 0.5\%, which is significantly lower than the specification of the vendors of 1.2\% and 2.6\%, respectively.
It follows that the particles are closer to monodispersity than declared, presumably since the specified distribution widths are estimates of the corresponding upper limit during production. 

The method for measurement and data analysis presented here allows to determine the mean diameter of an ensemble of \SI{2}{\micro \metre} PS spheres with an uncertainty of \SI{2}{nm} 
or a relative uncertainty of $10^{-3}$.
This uncertainty was estimated from the measurement noise and the limited precision of the inverse problem's solution and verified with synthetic data.
Model errors might have an additional effect not analyzed here, \eg,  deviations from a spherical shape or \emph{surface roughness}.
However, since even the most intricate details of the extinction spectra, \ie, their ripple structure, can be fitted with the model, it seems unlikely 
that surface roughness is relevant.
The scattering properties from ensembles of rough spheres have been examined \cite{Schiffer1989irregular, Schiffer1990perturbation}
and it was found that the deviations between spheres and rough spheres are largest for side scattering and backscattering amplitudes,
whereas forward and near-forward scattering is least sensitive to irregularities. 
Due to the   optical theorem \cite{BohrenHuffman}, this means that the extinction cross section $C_\text{ext}$ is
very insensitive to irregularities of the particles' surfaces. Hence the Mie scattering formulae may be used
even for somewhat irregular particles.

We have demonstrated the potential of spectral extinction measurements in combination with an adapted mathematical model to determine microparticle properties.
To further improve the model, in particular for new applications, the investigations will be extended to particles in the size range of several hundred nanometers 
up to about \SI{10}{\micro m} and to other (non-absorbing and absorbing) materials. In addition, comparison with complementary methods used to determine size and size distribution,
\ie, dynamic light scatter, secondary electron microscopy, nanoparticle tracking and sedimentation analyses will be carried out.

Potential applications of our spectral extinction measurement and analysis method include quality assurance to monitor size, size distribution and composition (\ie, spectral RI)
when producing suspensions of nano- and micro-particles or emulsions. The method is also sensitive towards the homogeneity of the particles (\eg, quartz spheres) 
and allows to identify irregularities by comparing the spectral RI of the particles to the RI of the bulk material. Complementarily, the roles of medium and particle
RI can be interchanged in the optimization. In this way, the spectral RI of the solvent, \eg, a protein solution, can be deduced using (monodisperse) micro-spheres with 
known optical and morphological parameters as a probe.

\appendix
\section{Expressions for the numerical  implementation of optimization}
\label{app:expressions}

The ensemble-averages with trapezoidal sums replacing integrals read
\begin{align}
 \overline{\mathcal{C}}(\lambda_i; \bm{\psi}) 
  &=   \sum_{t=1}^{I} \mathcal{C}(\lambda_i;  \mathfrak{n}_i(\bm{a}), R_t) \,\tilde{r}_t(\bm{\theta}).
\end{align}
For the Jacobian matrix one finds
\begin{align}
 J_{ij} = \frac{\partial F_i}{\partial \psi_j} &= 
 \begin{cases}
     [1+\epsilon(\lambda; \bm{\eta})]\frac{\partial \overline{\mathcal{C}}(\lambda_i; \bm{a}, \bm{\theta})}{\partial \psi_j} & j=1..L-M_\eta\\
     \frac{\partial \epsilon(\lambda; \bm{\eta})}{\partial \psi_j}\overline{\mathcal{C}}(\lambda_i; \bm{a}, \bm{\theta}) & \text{else}
 \end{cases},
\end{align}
with
\begin{align}
 \frac{\partial \overline{\mathcal{C}}(\lambda_i;\bm{a}, \bm{\theta})}{\partial a_j} 
 & = 
  \sum_{t=1}^{I} \left[\frac{\partial}{\partial \mathfrak{n}_i}\mathcal{C}(\lambda_i; \mathfrak{n}_i, R_{t})\right]\, \tilde{r}_{t}
 \,\frac{\partial \mathfrak{n}_i }{\partial a_j} 
 \nonumber\\
 &=
  \sum_{t=1}^{I} \left[\frac{\partial}{\partial \mathfrak{n}_i}\mathcal{C}(\lambda_i; \mathfrak{n}_i, R_{t})\right]\, \tilde{r}_{t}
 \,G_{i j}
 \\
 \frac{\partial \overline{\mathcal{C}}(\lambda_i;\bm{a}, \bm{\theta})}{\partial \theta_j} 
  &=
  \sum_{t=1}^{I} \mathcal{C}(\lambda_i;  \mathfrak{n}_i(\bm{a}), R_t)\,\frac{\partial}{\partial \theta_j}\tilde{r}_t(\bm{\theta})
\end{align}
A partial derivative like $\frac{\partial}{\partial \mathfrak{n}_i}\mathcal{C}(\lambda_i; \mathfrak{n}_i, R)$ can be computed numerically by finite differences at the same computational complexity
as $\mathcal{C}(\lambda_i; \mathfrak{n}_i, R)$. The derivatives $\frac{\partial \epsilon(\lambda; \bm{\eta})}{\partial \eta_j}$ are trivial because of the linear dependence of $\epsilon$ on $\bm{\eta}$.
The Jacobian matrix $\mathsf{J}$ is thus implemented efficiently using the above equations and was used explicitly in the numerical optimization.

\section{Uncertainty analysis}
\label{app:unc_analysis}

The sources of measurement uncertainties of $\overline{C}^*\!(\lambda)$ and estimates thereof were given in subsection~\ref{sec:exp_uncertainty}.
We now analyze the uncertainties of the resulting parameters obtained by nonlinear optimization. 
There are two types of uncertainty contributions:
\begin{enumerate}
 \item A contribution due to the measurement uncertainties of the experimental data $\overline{C}^*\!(\lambda)$. Here we focus on detector noise
 since concentration uncertainties and the compensation curve are included explicitly in the forward model.
 We will denote the terms related to the measurement uncertainty  by a superscript ``meas''.
 \item
 A contribution from the local least-squares optimization, denoted by a superscript ``init''.
 It describes the uncertainty  arising from the ambiguity of the solution  when starting the parameter
 optimization from different initial values.
\end{enumerate}
We estimated both contributions using the direct sampling Monte Carlo (MC) method as described in the following.

\subsection{Propagation of measurement uncertainties}
The contribution of measurement uncertainties is estimated by
linearized covariance-matrix propagation according to
\begin{equation}
 \mathsf{V}^\text{meas}(\bm{\psi}) =  \left[\mathsf{J}^T\, \mathsf{V}^\text{meas}(\bm{\overline{C}^*})^{-1}\,    \mathsf{J} \right]^{-1}.
 \label{eq:psi_lin_unc_estimate}
\end{equation}
To verify the validity of this linearization of the forward model, we performed MC simulations for synthetic data:
A number $N_{\rm MC}$ of synthetic datasets was created by adding independent random realizations of the measurement noise [Eq.~\eqref{eq:bathtub_model}]
to the same synthetic curve ($\mean(R)  = \SI{1}{\micro\metre}$, $\std(R) = \SI{5}{nm}$, $\epsilon\equiv0$). For each of these random datasets,
the inverse problem was solved numerically using the same initial parameter values. The covariance matrix of the optimized parameter vector $\bm{\psi}$ was estimated
for each of these numerical solutions according to Eq.~\eqref{eq:psi_lin_unc_estimate} and an average was taken over the $N_{\rm MC}$ samples.
In addition, we estimated $\mathsf{V}^\text{meas}(\bm{\psi})$ using
the statistical sample covariance over the set of MC samples. In this setup, the two independent estimates are consistent, thus legitimating the linearized
propagation of measurement uncertainties, also for experimental data with similar parameters.

\subsection{Uncertainties from initial values}
As described in subsection~\ref{sec:LSQ} 
the result for the parameter vector is
estimated as the weighted average from $N_\text{init}$ optimization runs with random initial values
\begin{equation}
 \langle\bm{\psi}\rangle_\text{init} := \frac{1}{Z_1} \sum_{i=1}^{N_\text{init}}  \beta_i \, \bm{\psi}^{(i)}
\end{equation}
with weights $\beta_i:= \exp\left[-\chi^2_{\rm dof}\left({\bm{\psi}}^{(i)}\right)\right]$ and $Z_s:=\sum_i {\beta_i}^s, s\in\mathbb{N}$.
$\bm{\psi}^{(i)}$ is the optimization result of the $i$th run.
From these statistical ensembles, we further estimated
the covariance
\begin{align}
\mathsf{V}^\text{init}(\langle\bm{\psi}\rangle_\text{init})& : =
\nonumber \\
&\hspace{-40pt}\frac{Z_2}{{Z_1}^2 - Z_2}\,\frac{1}{Z_1}  \sum_{i=1}^{N_\text{init}}  \beta_i \,
    \left( \bm{\psi}^{(i)} - \langle\bm{\psi}\rangle_\text{init}\right)
    \left( \bm{\psi}^{(i)} - \langle\bm{\psi}\rangle_\text{init}\right)^T  , 
\end{align}
which can be applied to both, experimental and synthetic data (keeping the realization of
measurement noise identical in all samples).
The total parameter variance corresponding to the combined uncertainty  of measurement and data analysis is then
\begin{equation}
\mathsf{V}(\langle\bm{\psi}\rangle_\text{init}) =  \left\langle\mathsf{V}^\text{meas}(\bm{\psi})\right\rangle_\text{init} + \mathsf{V}^\text{init}(\langle\bm{\psi}\rangle_\text{init}).
\end{equation}
For a sufficiently large number of samples, the ``initial'' term will ultimately scale like $ \mathsf{V}^\text{init}\propto1/\sqrt{ N_\text{init}}$.
Using $N_\text{init}=50$, the computing time was a few minutes on a year 2014 desktop PC and $ \mathsf{V}^\text{init}$ indeed became almost negligible.

From the parameter covariance $\mathsf{V}(\langle\bm{\psi}\rangle_\text{init})$ we can easily extract the uncertainty of the spectral refractive index $\bm{n} = \mathsf{G}\,\bm{a}$ as
\begin{equation}
 \mathsf{V}(\langle\bm{n}\rangle_\text{init}) = \mathsf{G}\,\mathsf{V}(\langle\bm{a}\rangle_\text{init})\, \mathsf{G}^T,
\end{equation}
where $\mathsf{V}(\langle\bm{a}\rangle_\text{init})$ corresponds to the first $M\times M$ entries of $\mathsf{V}(\langle\bm{\psi}\rangle_\text{init})$,
and analogously for the other resulting parameters.

\paragraph{Initial parameter values}
For analyzing \SI{2}{\micro\metre} PS particles, the  initial values were chosen as follows ($\mean\pm 1\std$ of Gaussian white noise):
\begin{enumerate}
 \item The RI was initialized as a piecewise-linear function spanned over the points $n[(260, 460, 800)\,\si{nm}]
 = (1.75, 1.605, 1.585)$, which is a crude approximation of the Sellmeier curve (Fig.~\ref{fig:nH2O_nPS}).
 For the MC sampling, the coefficient vector varied $\bm{a}$ with a standard deviation of 0.6 for $a_1$ and 0.04 for $a_j, j=2,...,M$, respectively.
 \item 
 The mean radius $\mean(R)$ was initialized to $\SI{1000}{nm}\pm\SI{20}{nm}$ and the distribution width $\std(R)$ to $\SI{10}{nm}\pm\SI{4}{nm}$.
 \item
 The compensation curve was initialized with $\eta_j = 0.00\pm 0.05$.
\end{enumerate}

\bigskip
\section*{Acknowledgements}
The authors thank Volker Ost and Uwe Sukowski for their support when designing the experiment, 
assembling the experimental setup and performing measurements.
We would also like to thank Sebastian Heidenreich  and Hermann Gro\ss{} for helpful discussions.

\bigskip
\bibliography{PS_RI_Cext_submission01}


\end{document}